\documentclass[conference]{IEEEtran}
\ifCLASSINFOpdf
\else
\fi

\usepackage{color}
\usepackage{amsmath,amssymb,amscd}
\usepackage{subfigure, epsfig}
\usepackage{pstricks}
\usepackage{pstricks-add}
\usepackage{pst-plot}

\newtheorem{lemma}{Lemma}
\newtheorem{theorem}{Theorem}
\newtheorem{definition}[theorem]{Definition}

\newcommand{\R}{\mathbb{R}}

\newcommand{\N}{\mathbb{N}}

\newcommand{\E}{\mathbb{E}}
\newcommand{\Q}{\mathbb{Q}}

\newcommand{\C}{\mathcal{C}}

\newcommand{\QQ}{\mathcal{Q}}
\newcommand{\M}{\mathcal{M}}

\newcommand{\PP}{\mathcal{P}}

\newcommand{\EE}{\mathcal{E}}
\newcommand{\mc}{\mathcal}

\begin{document}
%
\title{An Achievable Rate Region for the Broadcast Wiretap Channel with Asymmetric Side Information}


\author{\IEEEauthorblockN{Ma\"{e}l Le Treust\IEEEauthorrefmark{1},
Abdellatif Zaidi\IEEEauthorrefmark{2} and
Samson Lasaulce\IEEEauthorrefmark{1}}
\IEEEauthorblockA{\IEEEauthorrefmark{1}Laboratoire des Signaux et Syst\`{e}mes,
CNRS  - Universit\'{e} Paris-Sud 11 - Sup\'{e}lec,
91191, Gif-sur-Yvette Cedex, France\\
Email: \{letreust\},\{lasaulce\}@lss.supelec.fr}
\IEEEauthorblockA{\IEEEauthorrefmark{2}Institut Gaspard Monge,
Universit\'{e} Paris-Est Marne La Vall\'{e}e,
77454, Marne La Vall\'{e}e Cedex 2, France\\
Email: abdellatif.zaidi@univ-mlv.fr}}
\maketitle

\begin{abstract}
The communication scenario under consideration in this paper corresponds to a multiuser channel with
 side information and consists of a broadcast channel with two legitimate receivers and an eavesdropper.
  Mainly, the results obtained are as follows. First, an achievable rate region is provided for the
  (general) case of discrete-input discrete-output channels, generalizing existing results. Second,
  the obtained theorem is used to derive achievable transmission rates for two practical cases of
  Gaussian channels. It is shown that known perturbations can enlarge the rate region of broadcast
  wiretap channels with side information and having side information at the decoder as well can increase
  the secrecy rate of channels with side information. Third, we establish for the first time an explicit connection between
  multiuser channels and observation structures in dynamic games. In this respect, we show how
  to exploit the proved achievability theorem (discrete case) to derive a
  communication-compatible upper bound on the minmax level of a player.
\end{abstract}


%
\IEEEpeerreviewmaketitle

\section{Introduction}\label{introduction}

The notion of secrecy in communication systems has been widely
studied since 1949 and the publication of \cite{Shannon(secrecy)1949}
 by Shannon. He introduced a measure of secrecy for
communication systems called equivocation. The secrecy capacity of the general wiretap
 channel which consists of one transmitter, one legitimate receiver, and one
eavesdropper has been determined in \cite{Wyner(Wiretap)1975}. In
\cite{CsiszarKorner(BroadcastConf)78}, the authors
extended this result assuming that both the
legitimate receiver (to which the confidential message is intended)
and the eavesdropper have to decode a common message. Regarding broadcast channels, there are at least three other relevant works.
The authors of \cite{Steinberg2005a} investigate a broadcast
channel with side information or state at the encoder. In this model, the transition probability is controlled
by a sequence of i.i.d. parameters whose realizations are known non-causally and perfectly
by the encoder. They conclude that in the Gaussian case, there is no loss of rate of communication.
The authors of \cite{Bagherikaram2008} provide an
achievable rate region for the broadcast channel with two
legitimate receivers (each of them having to decode a private and
a confidential message) and an eavesdropper; the corresponding region
is shown to be tight in the case of physically degraded broadcast
channels. For the case of reversely degraded parallel broadcast
channels, one eavesdropper, and an arbitrary number of legitimate
receivers, the authors of \cite{KhistiTchamkertenWornell08}
determined the secrecy capacity for transmitting a common message,
and the secrecy sum-capacity for transmitting independent messages.

As far as the present work is concerned, the most relevant
contribution is provided in \cite{ChenVinck08}. Therein, the authors
provide an achievable rate of the discrete or general wiretap channel
when a side information is known non-causally to the transmitter (in
the sense of \cite{gelfand-it-1980}). Their
achievable secured rate is the minimum between
the secure rate of the wiretap channel \cite{Wyner(Wiretap)1975}
and the rate of the channel with side information provided by
Gel'fand and Pinsker in \cite{gelfand-it-1980}. The coding scheme in
\cite{ChenVinck08} is proved to achieve at least
one of these two rates and also satisfy the security
constraints $R\leq\frac{H(m|Z^n)}{n}$ where $m$ is the source message, $n$ is the codeword size, and $Z^n$ the observation vector of the eavesdropper.

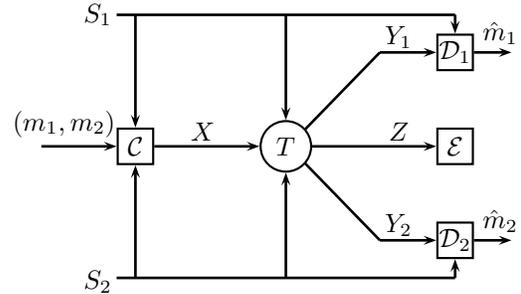
\begin{figure}[!ht]
\begin{center}
\psset{xunit=0.5cm,yunit=0.5cm}
\begin{pspicture}(-1,-4)(12,4)
\rput[u](1.5,0){$\mc{C}$}
\rput[u](10,2.5){$\mc{D}_1$}
\rput[u](10,-2.5){$\mc{D}_2$}
\rput[u](10,0){$\mc{E}$}
\rput[u](-0.4,0.6){$(m_1,m_2)$}
\rput[u](3.3,0.4){$X$}
 \rput[u](5.5,0){$T$}
\rput[u](8.5,2.9){$Y_1$}
\rput[u](8.5,-2.1){$Y_2$}
\rput[u](8.5,0.4){$Z$}
\rput[u](0.5,3.5){$S_1$}
 \rput[u](0.5,-3.6){$S_2$}
\rput[u](11.2,3){$\hat{m}_1$}
 \rput[u](11.2,-2){$\hat{m}_2$}
\psframe(1,-0.5)(2,0.5)
\psframe(9.5,2)(10.5,3)
\psframe(9.5,-3)(10.5,-2)
\psframe(9.5,-0.5)(10.5,0.5)
\pscircle(5.5,0){0.35}
\psline[linewidth=1pt]{->}(10.5,2.5)(11.5,2.5)
\psline[linewidth=1pt]{->}(10.5,-2.5)(11.5,-2.5)
\psline[linewidth=1pt]{->}(-1,0)(1,0)
\psline[linewidth=1pt]{->}(2,0)(4.8,0)
\psline[linewidth=1pt](6,0.44)(8,2.5)
\psline[linewidth=1pt](6,-0.44)(8,-2.5)
\psline[linewidth=1pt]{->}(8,2.5)(9.5,2.5)
\psline[linewidth=1pt]{->}(8,-2.5)(9.5,-2.5)
\psline[linewidth=1pt]{->}(1.5,-3.5)(1.5,-0.5)
\psline[linewidth=1pt]{->}(1.5,3.5)(1.5,0.5)
\psline[linewidth=1pt](1,-3.5)(10,-3.5)
\psline[linewidth=1pt](1,3.5)(10,3.5)
\psline[linewidth=1pt]{->}(5.5,-3.5)(5.5,-0.7)
\psline[linewidth=1pt]{->}(5.5,3.5)(5.5,0.7)
\psline[linewidth=1pt]{->}(10,3.5)(10,3)
\psline[linewidth=1pt]{->}(10,-3.5)(10,-3)
\psline[linewidth=1pt]{->}(6.2,0)(9.5,0)
\label{fig:BroadcastWSC}
\end{pspicture}
\caption{The broadcast wiretap channel with asymmetric side information.
The encoder $\mc{C}$ send the message $m_1$ (resp. $m_2$) to decoder $\mc{D}_1$ (resp. $\mc{D}_2$) through the channel $T$
by preventing the eavesdropper $\mc{E}$ to decode. $X$ is the channel input, $S_1$ and $Y_1$ (resp. $S_2$ and $Y_2$)
are the side information and the channel output available at the first (resp. second) decoder,
$Z$ is the channel output for the eavesdropper.}
\end{center}
\end{figure}

We extend this result by considering
the broadcast channel with confidential messages represented in Fig. 1. 
With respect to \cite{ChenVinck08}, two differences have to be noticed.
\begin{itemize}
\item A two-user broadcast channel is considered.
\item Each legitimate receiver only knows a part of the side information.
\end{itemize}
 To be more precise, if
$(S_1,S_2)$ represents the pair of side information, receiver or decoder $\mc{D}_k$, with $k\in \{1,2\}$, knows only
$S_k$. On the other hand, the eavesdropper $\mc{E}$ does not know the side information at all.

The two main motivations for deriving an achievable rate region for this multiuser channel are as follows. First of all,
the goal is to better understand the influence of the
side information on the performance limits of secure communications.
The second strong
motivation is more original since we show that coding theorems are also useful for understanding strategic interactions (games).
Indeed, as mentioned in \cite{Lasaulce-Tutorial-09}, there has been, in recent years, a surge of interest for game theory
since it can be useful to analyze multiuser settings (the interference channel is one of them \cite{YuGinisCioffi2002}, \cite{ScutariPalomarBarbarossa09}).
In those studies, quite often, Shannon transmission rates are considered for the player's utilities and game-theoretic notions
are applied. One of the messages of the present work is that, conversely, multiuser channels can be used to understand (dynamic)
games with arbitrary observation structures and utility functions. This contributes to strengthen the links between Shannon
theory and game theory and gives more momentum to some works in this direction such as \cite{AumannMashler(BookRGIncInfo)95}
\cite{RenaultTomala(GeneralProp)11}.

In the next section \ref{section:ChannelModel}, we introduce the channel model under investigation and the main achievability
 result (Theorem \ref{MainTheorem}). We compare the derived result with previous works in Sec. \ref{section:Interpretation}. In Sec.
  \ref{section:ProofTheorem}, we prove theorem \ref{MainTheorem}. Sec. \ref{section:GaussianChannel} is devoted to exploiting the derived theorem in the Gaussian case (achievability theorems follow in the Gaussian case provided long but simple calculations are done, the latter are omitted here).
We consider, in Sec. \ref{section:MinmaxLevels}, a direct application of our result to games. We provide an upper bound
on the min-max level in a four-player long-run game with a given observation/monitoring structure (called games with signals
in the literature of game theory). We conclude the paper by summarizing remarks and possible extensions of this work (Sec. \ref{conclusion}).

\section{Channel model}\label{section:ChannelModel}
In this paper, we denote $X,S_1,S_2,Y_1,Y_2,Z$ the random variables of the channel inputs $x\in \mc{X}$, the side information at the first $s_1\in \mc{S}_1$ and the second $s_2\in \mc{S}_2$ decoders, the channel ouputs for the first $y_1\in \mc{Y}_1$ and the second $y_2\in \mc{Y}_2$ decoders and the channel ouputs $z\in \mc{Z}$ for the eavesdropper (see Fig 1). The corresponding sequences will be written $Z^n=(Z(1),\ldots,Z(n))$, where superscripted letters denote the vector. The messages $m_1$ and $m_2$ are uniformly distributed among the sets $\mc{M}_1$ and $\mc{M}_2$ whose cardinalities are denoted $M_1=|\mc{M}_1|$ and $M_2=|\mc{M}_2|$. $\Delta(\mc{Y})$ denote the set of probability distributions over the set $\mc{Y}$, $\PP^{\otimes n}\in \Delta(\mc{X}^n)$ denote the $n$-times product of the probability $\PP\in \Delta(\mc{X})$ and $\text{co }\mc{R}$ denote the convex hull of a set $\mc{R}$.

Consider a broadcast wiretap channel with asymmetric side information, as a transition probability described in figure 1
\begin{eqnarray}
T : \mc{X} \times \mc{S}_1 \times\mc{S}_2 \longrightarrow \Delta(\mc{Y}_1 \times\mc{Y}_2 \times\mc{Z}).
\end{eqnarray}
The side information $s_1,s_2$ are drawn independently and identically distributed from the joint distribution $P_s\in \Delta(\mc{S}_1\times \mc{S}_2)$. The sequence of realizations $s_1^n,s_2^n$ are non-causally known at the encoder and at their respective decoders. The channel is discrete and memoryless, i.e.  the $n$-stage transition probability is defined as follows:
\begin{eqnarray}
&&T^{\otimes n}(y_1^n,y_2^n,z^n|x^n,s_1^n,s_2^n) \nonumber\\
&&= \prod_{i=1}^n T(y_{1}(i),y_{2}(i),z(i)|x(i),s_{1}(i),s_{2}(i)) .  
\end{eqnarray}
\begin{definition}
Define an $(n,M_1,M_2)$-code as a triplet of functions as follows:
\begin{eqnarray}
&f& : \mc{M}_1 \times \mc{M}_2 \times \mc{S}_1^n\times  \mc{S}_2^n \longrightarrow \mc{X}^n,\\
&g_1& : \mc{Y}_1^n\times \mc{S}_1^n \longrightarrow {\mc{M}}_1, \\
&g_2& : \mc{Y}_2^n \times \mc{S}_2^n\longrightarrow {\mc{M}}_2.
\end{eqnarray}
$(\hat{m}_1,\hat{m}_2)$ denote the random variable of the messages reconstructed by the code.
Define the error probability $\PP_e^n$ associated with each $(n,M_1,M_2)$-code as follows:
\begin{eqnarray}
\PP_e^n = \PP((m_1,m_2)\neq(\hat{m}_1,\hat{m}_2)).
\end{eqnarray}
\end{definition}

The amount of information of a code is related to the cardinality $M_1$ and $M_2$ of the sets of messages $\mc{M}_1$ and $\mc{M}_2$. As in \cite{shannon-bell-1948}, this quantity is measured by the rate $R=\frac{\log M}{n}$ of the code. In the context of secure communication, the notion of equivocation $\frac{H(m|Z^n)}{n}$ \cite{Shannon(secrecy)1949} is introduced as a measure of the secrecy level guaranteed by a code. When this level is greater than the rate of the code, it prevents the eavesdropper from correctly decoding the transmitted information.

\begin{definition}
A rate pair $(R_1,R_2)$ is said to be achievable if for all $\varepsilon>0$, there exists a $(n,M_1,M_2)$-code such that:
\begin{eqnarray}
\frac{\log M_1}{n} &\geq& R_1-\varepsilon,\\
\frac{\log M_2}{n} &\geq& R_2-\varepsilon,\\
\frac{H(m_1|Z^n)}{n} &\geq& R_1-\varepsilon,\\
\frac{H(m_2|Z^n)}{n} &\geq& R_2-\varepsilon,\\
\frac{H(m_1,m_2|Z^n)}{n} &\geq& R_1+R_2-\varepsilon,\\
 \PP_e^n &\leq& \varepsilon.
\end{eqnarray}
Denote $\mc{R}$ the set of achievable rate pairs.
\end{definition}

\subsection{Main result}
We provide an achievable rate region for the considered broadcast wiretap channel with asymmetric side information.
\begin{definition}
Denote $\mc{R}_I$ the set of rate pairs $(R_1,R_2)$ such that there exists a probability distribution $\PP(u_1,u_2,x|s_1,s_2)$ satisfying:
\begin{eqnarray}
R_1&\leq & I(U_1;Y_1,S_1) - \max(I(U_1;Z), I(U_1;S_1,S_2)), \nonumber\\
R_2&\leq & I(U_2;Y_2,S_2) - \max(I(U_2;Z), I(U_2;S_1,S_2)),  \nonumber \\
R_1+R_2 &\leq & I(U_1;Y_1,S_1) +I(U_2;Y_2,S_2)  - I(U_1;U_2) \nonumber\\
&& - \max(I(U_1,U_2;Z), I(U_1,U_2;S_1,S_2) ).\label{inequalities:maintheorem}
\end{eqnarray}
\textit{Remark} that the probability $\PP(u_1,u_2,x|s_1,s_2)$ induces a general distribution $\QQ$ that satisfies the Markov
property $(U_1,U_2)-(X,S_1,S_2) - (Y_1,Y_2,Z)$. This probability $\QQ$ is defined for every $(u_1,u_2,x,s_1,s_2,y_1,y_2,z)$, by the following equation:
\begin{eqnarray*}
&&\QQ(u_1,u_2,x,s_1,s_2,y_1,y_2,z) = \\
&&P_s(s_1,s_2)\times \PP(u_1,u_2,x|s_1,s_2)\times T(y_1,y_2,z|x,s_1,s_2).
\end{eqnarray*}
\end{definition}
\begin{theorem}\label{MainTheorem}
Any rate pair $(R_1,R_2)\in \text{co } \mc{R}_I $ is achievable for the broadcast wiretap channel with asymmetric side information.
\end{theorem}


Suppose that we need the channel input to be correlated with a sequence of i.i.d. random variable $S^n$.
The analysis leads to consider the random variable $S$ as a side information even if it does not impact the transition probability.
This remark applies, more specifically, in a game theoretical framework (see Sec. \ref{section:MinmaxLevels}).

\section{Interpretation}\label{section:Interpretation}

The achievable rate region $\mc{R}_I$ we provide is a generalization of the one in \cite{ChenVinck08}. It consists in the intersection of two rate regions. The first one is related to the side information as in \cite{Steinberg2005a} and the second one is related to the eavesdropper as in \cite{Bagherikaram2008}.
Note that if we remove the eavesdropper ($Z=\mc{C}$) and we consider that the side information is non-causally known only at the encoder, our rate region boils down to the one of  \cite{Steinberg2005a} when the variable $W$ is constant. If we remove the side information ($S_1=S_2=\mc{C}$), the rate region equals the one described in \cite{Bagherikaram2008}.
Suppose we remove  the receivers ($\mc{D}_2$) and the  side information ($S_1=\mc{C}$) and in that case the rate region boils down to the one of the article \cite{ChenVinck08}.

\section{Proof of theorem \ref{MainTheorem}}\label{section:ProofTheorem}
We first prove the achievability of the rate pair $(R_1,R_2)\in \mc{T}_I $ satisfying the above inequalities (\ref{inequalities:maintheorem}). Fix a distribution $\QQ(u_1,u_2,x,s_1,s_2,y_1,y_2,z)$ satisfying the channel transition $T(y_1,y_2,z|x,s_1,s_2)$, the distribution $P_s(s_1,s_2)$ and the rates inequalities (\ref{inequalities:maintheorem}). We will prove that the pair $(R_1,R_2)\in \mc{T}_I $ is achievable. Denote $A_{\varepsilon}^{*{n}}(U_1\times U_2 |s_1^n,s_2^n)$ the set of sequences
$u_1^n,u^n_2$ that are jointly typical with $s_1^n,s_2^n$. The properties of the typical sequences can be founded in \cite{Cover-Book-91} and \cite{CsiszarKorner(Book)81}.
\begin{itemize}
\item Generation of the Code-book : Generate $M_{Y_1}=2^{n R_{Y_1}}=2^{n(I(U_1;Y_1,S_1)-\varepsilon)}$
 sequences $u_1^{n}$ from distribution $\QQ_{U_1}(u_1)^{\otimes n}$. Distribute them at random into
 $M_1=2^{nR_1}$ bins denoted $i_1\in \{1,\ldots,M_1\}$, containing each of them $M_{U_1}=2^{nR_{U_1}}$
 sequences $u_1^{n}$. Divide each bin  $i_1$ into $M_{W_1}=2^{nR_{W_1}}$
    sub-bins denoted $j_1\in \{1,\ldots,M_{W_1}\}$ containing each of them $M_{Z_1}=2^{nR_{Z_1}}$
    sequences $u_1^{n}$ with the following parameters $R_{U_1}, R_{Y_1}, R_{1}, R_{Z_1}$.
    Generate $M_{Y_2} =2^{n  R_{Y_2} }=2^{n(I(U_2;Y_2,S_2)-\varepsilon)}$ sequences $u_2^{n}$ from
    distribution $\QQ_{U_2}(u_2)^{\otimes n}$. Distribute them at random into $M_2 =2^{n R_2 }$ bins
    denoted $i_2\in \{1,\ldots,M_2 \}$, containing each of them $M_{U_2} =2^{nR_{U_2} }$ sequences
    $u_2^{n}$. Divide each bin  $i_2$ into $M_{W_2}=2^{nR_{W_2}}$
    sub-bins denoted $j_2 \in \{1,\ldots,M_{W_2} \}$ containing each of them $M_{Z_2}=2^{nR_{Z_2}}$
    sequences $u_2^{n}$ with the above parameters $R_{U_2} , R_{Y_2} , R_{2} , R_{Z_2}$.
    For each tuple of sequences $(u_1^n,u_2^n,s_1^n,s_2^n)$ draw a sequence $x^n$ from the
    distribution $\QQ(x|u_1,u_2,s_1,s_2)^{\otimes n}$.
\item Encoder obtains the message $(i_1,i_2)\in \M_1 \times \M_2$ and the sequence of side information
$(s_1^n,s_2^n)$. It finds a pair of sequences $u_1^{n}$ in the bin $i_1 $ and $u_2^n$ in the bin
$i_2 $ such that $(u_1^n,u_2^n)\in A_{\varepsilon}^{*{n}}(U_1\times U_2 |s_1^n,s_2^n)$. Send the
sequence $x^n$ corresponding to the tuple of sequences $(u_1^n,u_2^n,s_1^n,s_2^n)$.
    \begin{eqnarray*}
R_{U_1} &>& I(U_1;S_2,S_1),\\
R_{U_2} &>& I(U_2;S_1,S_2),\\
R_{U_1}  +  R_{U_2} &>& I(U_1;U_2) + I(U_1,U_2;S_1,S_2),\\
&&\\
R_{Y_1}  = R_{U_1}  + R_{1}  &<& I(U_1;Y_1,S_1),\\
R_{Y_2}  = R_{U_2}  + R_{1}  &<& I(U_2;Y_2,S_2),\\
&&\\
R_{Z_1}&<& I(U_1;Z),\\
R_{Z_2}&<& I(U_2;Z),\\
R_{Z_1} +  R_{Z_2}&<& I(U_1;U_2)+ I(U_1,U_2;Z),\\
&&\\
R_{U_1} &>& R_{Z_1},\\
R_{U_2} &>& R_{Z_2}.
\end{eqnarray*}
\item Decoder 1 receives the channel output $y_1^n$ and the sequence of side information $s_1^n$. It finds a unique sequence $u_1^n$ such that $u_1^n\in A_{\varepsilon}^{*{n}}(U_1|y_1^n, s_1^n)$ and it returns the bin index $i_1$ of the sequence $u_1^n$.
\item Decoder 2 receives the channel output $y_2^n$ and the sequence of side information $s_2^n$. It finds a unique sequence $u_2^n$ such that $u_2^n\in A_{\varepsilon}^{*{n}}(U_2|y_2^n, s_2^n)$ and it returns the bin index $i_2$ of the sequence $u_2^n$.
\end{itemize}

The proof consists first to show that the error probability can be upper bounded by $\varepsilon>0$ as $n$ goes to infinity. Second, we check if the equivocation rate at the eavesdropper is sufficiently high as $n$ goes to infinity. We conclude that the desired rate $(R_1,R_2)$ pair belongs to the achievable rate region that satisfies by the above inequalities (\ref{inequalities:maintheorem}).

\textbf{Analysis of the error probability.} As in the articles \cite{gelfand-it-1980} and \cite{marton-it-1979}, it is based on extensions of the following lemma:
\begin{lemma}\label{lemma:TypicalSequences}
\textit{The properties of the typical sequences \cite{CsiszarKorner(Book)81}}.
Let the joint probability $\QQ(x,y)\in \Delta(X\times Y)$, then:
\begin{eqnarray*}
\QQ^{\otimes n}(x^n\in A_{\varepsilon}^{n*}(X|y^n)|y^n)\geq 1-\varepsilon\qquad \forall y^n\in A_{\varepsilon}^{n*}(Y)\label{eq:CondTypicity}.
\end{eqnarray*}
\end{lemma}
\begin{lemma}\label{lemma:MutualProb} \textit{The mutual covering lemma \cite{elgamal-it-1981}}.
Suppose that the family of sequences $(u(i)^n)_{i\in 2^{nR_I}}\in U^n$ is drawn i.i.d. from $\QQ_U^{\otimes n}$ and $(v(j)^n)_{j\in 2^{nR_J}}$ is drawn i.i.d. from $\QQ_V^{\otimes n}$. Then for all $\varepsilon>0$, there exists an $\bar{n}\geq 0$ such that for all $n\geq\bar{n}$:
\begin{eqnarray*}
&&R_I +R_J < I(U;V) \Longrightarrow\\
&&\PP(\cup_{i\in I,\atop j\in J}\{(u(i)^n,v(j)^n)\in A_{\varepsilon}^{*n}(U\times V)\} )\leq \varepsilon, \label{eq:mutualcovering1}\\
&&R_I +R_J > I(U;V) \Longrightarrow\\
&&  \PP(\cap_{i\in I,\atop j\in J}\{(u(i)^n,v(j)^n)\notin A_{\varepsilon}^{*n}(U\times V)\} )\leq \varepsilon.
\label{eq:mutualcovering2}
\end{eqnarray*}
\end{lemma}
Without loss of generality, we assume that the encoder has to transmit the messages $(i_1,i_2)$. Denote $B_{i_1} $ and $B_{i_2} $ the bins of sequences $u_1^n$ and $u_2^n$ respectively. Let us define the following error events:
\begin{itemize}
\item $\EE_1=\{ (s_1^n,s_2^n)\notin A_{\varepsilon}^{*{n}}( S_1\times S_2)\}$ the two sequences of side information are not jointly typical.
\item $\EE_2=\{\forall (u_1^n,u_2^n) \in B_{i_1}\times B_{i_2} ,\; (u_1^n,u_2^n)\notin A_{\varepsilon}^{*{n}}(U_1 \times U_2 |s_1^n,s_2^n)\}$ there is no pair of sequence $(u_1^n,u_2^n) $ in the bins $B_{i_1}$ and $B_{i_2}$ that are jointly typical with $(s_1^n,s_2^n)$.
\item $\EE_3=\{(x^n,y_1^n,y_2^n,z^n) \notin A_{\varepsilon}^{*{n}}( X\times Y_1\times Y_2\times Z|u_1^n,u_2^n,s_1^n,s_2^n)| (u_1^n,u_2^n,s_1^n,s_2^n) \in A_{\varepsilon}^{*{n}}( U_1 \times U_2 \times S_1\times S_2\}$ the family $(x^n,y_1^n,y_2^n,z^n)$ of sequences is not jointly typical with the jointly typical sequences $(u_1^n,u_2^n,s_1^n,s_2^n)$.
\item $\EE_4 =\{\exists u_1^{'n} \neq u_1^n, \; (u_1^{'n},y_1^n,s_1^n)^n\in A_{\varepsilon}^{*{n}}(U_1\times Y_1\times S_1)\}$ there is another vector $u_1^{'n}$ jointly typical with the channel output $y_1^n$ and the side information $s_1^n$.
\item $\EE_5 =\{\exists u_2^{'n} \neq u_2^{n}, \; (u_2^{'n},y_2^{n},s_2^{n})\in A_{\varepsilon}^{*{n}}(U_2\times Y_2 \times S_2)\}$ there is  another vector $u_2^{'n}$ jointly typical with the channel output $y_2^n$ and the side information $s_2^n$.
\end{itemize}
Using an extension of covering lemma \cite{elgamal-it-1981}, we bound $\PP(\EE_2)$ by $\varepsilon$ as soon as, the following inequalities are satisfied.
\begin{eqnarray}
R_{U_1} &>& I(U_1;S_2,S_1)\label{eq:RateConstraintsU1},\\
R_{U_2} &>& I(U_2;S_1,S_2)\label{eq:RateConstraintsU2},\\
R_{U_1}  +  R_{U_2} &>& I(U_1;U_2) +I(U_1,U_2;S_1,S_2). \label{eq:RateConstraintsU12}
\end{eqnarray}
$\PP(\EE_4)$ and $\PP(\EE_5)$ are bounded by $\varepsilon$ if:
\begin{eqnarray}
R_{Y_1}  = R_{U_1}  + R_{1}  &<& I(U_1;Y_1,S_1),\label{eq:RateConstraintsY1}\\
R_{Y_2}  = R_{U_2}  + R_{1}  &<& I(U_2;Y_2,S_2).\label{eq:RateConstraintsY2}
\end{eqnarray}
To bound $\PP(\EE_1)$ and $\PP(\EE_3)$, we use classical properties of the typical sequences \cite{CsiszarKorner(Book)81}.
Thus for all $\varepsilon$, there exists $n$ such that,
\begin{eqnarray}
\PP_e^n\leq 5\varepsilon.
\end{eqnarray}
We proved that the error probability is upper bounded by $5\varepsilon$.

 \textbf{The equivocation rate at the eavesdropper.} \\
 Denote $(m_1,m_2)$ the random variable of the pair of bins and $(w_1,w_2)$ the random variable of the pair of sub-bins. Let us prove that $\frac{H(m_1,m_2|Z^n)}{n} \geq R_1+R_2-\varepsilon$. We first introduce the random variables $w_1,w_2$ and $U_1^n,U_2^n$ in the expression of $H(m_1,m_2|Z^n)$.
\begin{eqnarray}
&&H(m_1,m_2|Z^n)   \nonumber  \\
&=& H(m_1,m_2,Z^n)-H(Z^n)   \nonumber  \\
&=& H(m_1,m_2,w_1,w_2,Z^n)   \nonumber  \\
&&-H(w_1,w_2|m_1,m_2,Z^n)-H(Z^n)   \nonumber  \\
&=& H(m_1,m_2,w_1,w_2,U_1^n,U_2^n, Z^n)   \nonumber  \\
&&-  H(U_1^n,U_2^n|m_1,m_2,w_1,w_2, Z^n)   \nonumber  \\
&& - H(w_1,w_2|m_1,m_2,Z^n)-H(Z^n)   \nonumber  \\
\end{eqnarray}
\begin{eqnarray}
&=& H(m_1,m_2,w_1,w_2|U_1^n,U_2^n, Z^n)   \nonumber  \\
&& +  H(U_1^n,U_2^n, Z^n)   \nonumber  \\
&& -  H(U_1^n,U_2^n|m_1,m_2,w_1,w_2, Z^n)   \nonumber  \\
&& - H(w_1,w_2|m_1,m_2,Z^n)-H(Z^n)   \nonumber  \\
&=& H(m_1,m_2,w_1,w_2|U_1^n,U_2^n, Z^n)\label{eq:equivocation1term}\\
&& +  H(U_1^n,U_2^n| Z^n)\label{eq:equivocation2term}\\
&& -  H(U_1^n,U_2^n|m_1,m_2,w_1,w_2, Z^n)\label{eq:equivocation3term}\\
&& - H(w_1,w_2|m_1,m_2,Z^n).\label{eq:equivocation4term}
\end{eqnarray}
We provide a lower bound for each of the four terms of the above equation.\\
\textit{The first term} (\ref{eq:equivocation1term}) in the above equation is removed.\\
\textit{The second term} (\ref{eq:equivocation2term}) is lower bounded, using the chain rule \cite{Cover-Book-91}, by the following quantity:
\begin{eqnarray*}
&&H(U_1^n,U_2^n| Z^n)  \\
&=&  H(U_1^n) +H(U_2^n) -I(U_1^n;U_2^n) - I(U_1^n,U_2^n;Z)\\
&\geq&  I(U_1^n;Y_1^n,S_1^n) +  I(U_2^n;Y_2^n,S_2^n)\\
&& -I(U_1^n;U_2^n)  - I(U_1^n,U_2^n;Z)  \\
&\geq&  n[I(U_1;Y_1,S_1) +  I(U_2;Y_2,S_2) \\
&&-I(U_1;U_2)  - I(U_1,U_2;Z)].
\end{eqnarray*}
\textit{The third term} (\ref{eq:equivocation3term}) is lower bounded by $-2\varepsilon -n 2\varepsilon \log |Z|$ using Fano's inequality \cite{Cover-Book-91} and the following system of conditions:
\begin{eqnarray}
R_{Z_1}&<& I(U_1;Z),\label{eq:RateConstraintsZ1}\\
R_{Z_2}&<& I(U_2;Z),\label{eq:RateConstraintsZ2}\\
R_{Z_1} +  R_{Z_2}&<& I(U_1;U_2)+ I(U_1,U_2;Z)\label{eq:RateConstraintsZ12}.
\end{eqnarray}
Denote $B_{m_1}$ the bin with index $m_1$ and $B_{w_1}$ the sub-bin with index $w_1$. Let the following events:
\begin{eqnarray*}
\EE_6 &=& \{\forall (u_1 ,u_2 )\in (B_{m_1} \times B_{m_2}  )\cap (B_{w_1}\times B_{w_2} ),\; \\
&&(u_1^n ,u_2^n ,z^n)\notin A_{\varepsilon}^{*{n}}(U_1\times U_2\times Z)\},\\
\EE_7 &=& \{\exists (u_1 ,u_2 )'\neq (u_1 ,u_2 )\\
&&\in (B_{m_1} \times B_{m_2}  )\cap (B_{w_1}\times B_{w_2} ),\\
&&\;s.t. (u'_1,u'_2,z)\in A_{\varepsilon}^{*{n}}(U_1\times U_2\times Z)\}.
\end{eqnarray*}
Consider a typical decoding function of the eavesdropper knowing the pairs of bin indexes $(m_1,m_2)$ and sub-bin index $(w_1,w_2)$,
\begin{eqnarray}
g : \mathcal{Z}^n \longrightarrow \mathcal{U}_1^n \times \mathcal{U}_2^n.
\end{eqnarray}
To the received sequence $z^n$, it associates the pair $(u_1^n,u_2^n)$ if it belong to the bins $(m_1,m_2)$, the sub-bins $(w_1,w_2)$ and is jointly typical with $z^n$. Define the error probability of such a decoding function
\begin{eqnarray}
\PP_{{\ae}}&=& \PP((U_1^n,U_2^n)\neq g(Z^n)\text{ s.t. } (U_1^n,U_2^n)\in\nonumber\\
&& (B_{m_1}\times B_{m_2} )\cap (B_{w_1}\times B_{w_2} ))\\
&\leq & \PP(\EE_6) + \PP(\EE_7) \leq 2\varepsilon,
\end{eqnarray}
where $\PP(\EE_6)\leq \varepsilon$ comes from properties of the typical sequences \cite{CsiszarKorner(Book)81} and  $\PP(\EE_7)\leq \varepsilon$ comes from the above system of equations (\ref{eq:RateConstraintsZ1})-(\ref{eq:RateConstraintsZ12}). Using Fano's inequality \cite{Cover-Book-91} we have:
\begin{eqnarray*}
&&H(U_1^n,U_2^n|m_1,m_2,w_1,w_2, Z^n)\\
&\leq& H(\PP_{{\ae}}) + n \PP_{{\ae}} (\log|Z|-\varepsilon)\\
&\leq& 2\varepsilon +n 2\varepsilon \log |Z|.
\end{eqnarray*}
\textit{The fourth term} (\ref{eq:equivocation4term}) is lower bounded by the following quantity:
$- n(\max[I(U_1,U_2;S_1,S_2) - I(U_1,U_2;Z), 0] +4 \varepsilon)$.
From the condition (\ref{eq:RateConstraintsU12}) and the definition of the sub-bins we have:
\begin{eqnarray}
R_{U_1}   + R_{U_2} &\geq& \max [ I(U_1,U_2;S_1,S_2)\nonumber\\
&& + I(U_1;U_2),  R_{Z_1}   + R_{Z_2}].
\end{eqnarray}
Suppose that the two following conditions are satisfied:
\begin{eqnarray}
R_{U_1}  + R_{U_2}   &\leq&  \max [I(U_1,U_2;S_1,S_2) \nonumber \\
&& + I(U_1;U_2),  R_{Z_1}   + R_{Z_2}] + 2 \varepsilon,\\
R_{Z_1}  + R_{Z_2}   &\geq&  I(U_1;U_2)+ I(U_1,U_2;Z) - 2\varepsilon.
\end{eqnarray}
We now prove the following inequalities:
\begin{eqnarray*}
&&H(w_1,w_2|m_1,m_2,Z^n)\\
&\leq& \log(|W_1 |\times |W_2 |)\\
&=& n (R_{U_1}   + R_{U_2}  -  R_{Z_1}   - R_{Z_2}  ) \\
&\leq& n (\max[ I(U_1;U_2) + I(U_1,U_2;S_1,S_2),\\
&&  I(U_1;U_2)  I(U_1,U_2;Z)] + 2\varepsilon \\
&& -  I(U_1;U_2) - I(U_1,U_2;Z) + 2\varepsilon ) \\
&\leq& n (\max[ I(U_1,U_2;S_1,S_2) - I(U_1,U_2;Z),   0] + 4\varepsilon).
\end{eqnarray*}
Combining the four above terms, we obtain the lower bound $R_1+R_2 - \bar{\varepsilon}$ over the equivocation rate.
\begin{eqnarray*}
&&H(m_1,m_2|Z^n) \\
&\geq &  n(I(U_1 ;Y_1 ,S_1 ) + I(U_2 ;Y_2 ,S_2 )\\
&&- I(U_1 ;U_2 ) - I(U_1 ,U_2 ;Z)) -2\varepsilon -n 2\varepsilon \log |Z|\\
&&- n (\max( I(U_1,U_2;S_1,S_2) - I(U_1,U_2;Z),   0) + 4\varepsilon) \\
&\geq &  n(I(U_1 ;Y_1 ,S_1 ) + I(U_2 ;Y_2 ,S_2 )\\
&&- I(U_1 ;U_2 )  - \max( I(U_1,U_2;S_1,S_2),I(U_1,U_2;Z))) \\
&&-  2\varepsilon -  n\varepsilon (2 \log |Z| +4)\\
&\geq & n (R_1+R_2) -  2\varepsilon -  n\varepsilon (2 \log |Z| +4).
\end{eqnarray*}
\vspace{-0.4cm}
\begin{eqnarray*}
\Longleftrightarrow  \frac{I(m_1,m_2;Z^n)}{n} & \leq& \bar{\varepsilon}.
\end{eqnarray*}
With $\bar{\varepsilon}= \varepsilon (2/n+2\log |Z|+4)$.
The same arguments apply to prove that:
\begin{eqnarray}
&&\frac{H(m_1|Z^n)}{n} \geq R_1 - \bar{\varepsilon},\\
&&\frac{H(m_2|Z^n)}{n} \geq R_2 - \bar{\varepsilon}.
\end{eqnarray}

\textbf{The transmission rates}
are determined by the binning scheme:
\begin{eqnarray*}
R_1 &=& R_{Y_1} - R_{U_1}, \\
R_2 &=& R_{Y_2} - R_{U_2},\\
R_1+R_2 &=&  (R_{Y_1}+ R_{Y_2}) -  R_{U_1} - R_{U_2}.
\end{eqnarray*}
We have proven that our coding scheme achieves every rate pair of the following rate region $ \mc{R}_I $.
\begin{eqnarray*}
R_1 &\leq& I(U_1;Y_1,S_1) - \max[I(U_1;Z) ,I(U_1;S_1,S_2)],\\
R_2 &\leq& I(U_2;Y_2,S_2) - \max[I(U_2;Z) ,I(U_2;S_1,S_2)], \\
R_1+R_2 &\leq&  I(U_1;Y_1,S_1) + I(U_2;Y_2,S_2) -I(U_1;U_2)\\
 &-& \max[I(U_1,U_2;Z) ;I(U_1,U_2;S_1,S_2)].
\end{eqnarray*}
A classical time-sharing argument in the coding scheme implies that the convex hull $ \text{co }\mc{R}_I $ of the rate region is achievable.

\section{The case of Gaussian channels}\label{section:GaussianChannel}

In this section, we want to show theorem \ref{MainTheorem} can be exploited for Gaussian communication channels. At least two interesting results are emphasized. For the first model under consideration (Fig. 2), it is shown that the presence of  known perturbations (namely $S_1$ and $S_2$) can enhance the secrecy rates. In fact, if those perturbations are sufficiently strong, it is even possible to obtain the same rate region as if the eavesdropper were not present. For the second model (Fig. 4), it is shown that knowing the side information can lead to a larger secrecy rate, which is usually not the case in channels with states but with no eavesdropper.

\subsection{Increasing the influence of known perturbations enhances the rate region}\label{subsection:CompensateEavesdropper}

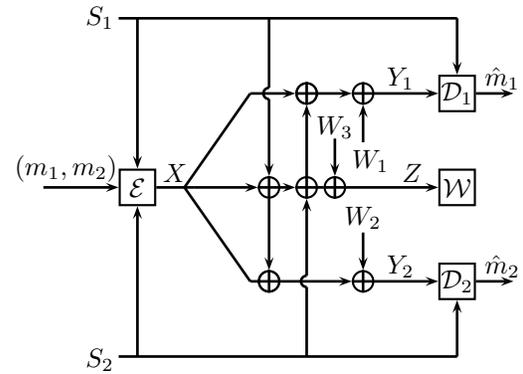
\begin{figure}[!ht]
\begin{center}
\psset{xunit=0.5cm,yunit=0.5cm}
\begin{pspicture}(-1,-5)(12,5)
\rput[u](1.5,0){$\mc{E}$}
\rput[u](10,2.5){$\mc{D}_1$}
\rput[u](10,-2.5){$\mc{D}_2$}
\rput[u](10,0){$\mc{W}$}
\rput[u](-0.4,0.5){$(m_1,m_2)$}
\rput[u](2.5,0.4){$X$}
\rput[u](8.5,2.9){$Y_1$}
\rput[u](8.5,-2.1){$Y_2$}
\rput[u](8.8,0.4){$Z$}
\rput[u](0.5,4.5){$S_1$}
\rput[u](0.5,-4.6){$S_2$}
\rput[u](11.2,2.9){$\hat{m}_1$}
\rput[u](11.2,-2.1){$\hat{m}_2$}
\rput[d](7.7,0.7){$W_1$}
\rput[u](6.75,1.6){$W_3$}
\rput[u](7.5,-0.8){$W_2$}
\psframe(1,-0.5)(2,0.5)
\psframe(9.5,2)(10.5,3)
\psframe(9.5,-3)(10.5,-2)
\psframe(9.5,-0.5)(10.5,0.5)
\pscircle(6,2.5){0.15}
\pscircle(7.5,2.5){0.15}
\pscircle(5,-2.5){0.15}
\pscircle(7.5,-2.5){0.15}
\pscircle(6.75,0){0.15}
\pscircle(5,-2.5){0.15}
\pscircle(6,0){0.15}
\pscircle(6,2.5){0.15}
\pscircle(5,0){0.15}
\psarc[linecolor=black,linewidth=1pt](6,-2.5){0.075}{90}{270}
\psarc[linecolor=black,linewidth=1pt](5,2.5){0.075}{90}{270}
\psline[linewidth=1pt]{->}(10.5,2.5)(11.5,2.5)
\psline[linewidth=1pt]{->}(10.5,-2.5)(11.5,-2.5)
\psline[linewidth=1pt]{->}(-1,0)(1,0)
\psline[linewidth=1pt](2,0)(3,0)
\psline[linewidth=1pt](2.75,0)(4.5,2.5)
\psline[linewidth=1pt](2.75,0)(4.5,-2.5)
\psline[linewidth=1pt]{->}(4.5,2.5)(5.7,2.5)
\psline[linewidth=1pt]{->}(5.7,2.5)(7.2,2.5)
\psline[linewidth=1pt]{->}(7.2,2.5)(9.5,2.5)
\psline[linewidth=1pt](4.5,-2.5)(4.7,-2.5)
\psline[linewidth=1pt]{->}(4.7,-2.5)(7.2,-2.5)
\psline[linewidth=1pt]{->}(7.2,-2.5)(9.5,-2.5)
\psline[linewidth=1pt]{->}(1.5,-4.5)(1.5,-0.5)
\psline[linewidth=1pt]{->}(1.5,4.5)(1.5,0.5)
\psline[linewidth=1pt](1,-4.5)(10,-4.5)
\psline[linewidth=1pt](1,4.5)(10,4.5)
\psline[linewidth=1pt](6,-4.5)(6,-2.65)
\psline[linewidth=1pt]{->}(6,-2.35)(6,-0.3)
\psline[linewidth=1pt](6,-0.3)(6,0.3)
\psline[linewidth=1pt]{->}(6,0.3)(6,2.2)
\psline[linewidth=1pt](5,4.5)(5,2.65)
\psline[linewidth=1pt]{->}(5,2.35)(5,0.3)
\psline[linewidth=1pt](5,0.3)(5,-0.3)
\psline[linewidth=1pt]{->}(5,-0.3)(5,-2.2)
\psline[linewidth=1pt](5,-2.2)(5,-2.8)
\psline[linewidth=1pt](6,2.2)(6,2.8)
\psline[linewidth=1pt]{->}(10,4.5)(10,3)
\psline[linewidth=1pt]{->}(10,-4.5)(10,-3)
\psline[linewidth=1pt]{->}(3,0)(4.7,0)
\psline[linewidth=1pt]{->}(4.7,0)(5.7,0)
\psline[linewidth=1pt](5.7,0)(6.45,0)
\psline[linewidth=1pt]{->}(6.45,0)(9.5,0)
\psline[linewidth=1pt]{->}(7.5,-1.2)(7.5,-2.2)
\psline[linewidth=1pt]{->}(6.75,1.3)(6.75,0.3)
\psline[linewidth=1pt]{->}(7.5,1.2)(7.5,2.2)
\psline[linewidth=1pt](7.5,-2.8)(7.5,-2.2)
\psline[linewidth=1pt](6.75,-0.3)(6.75,0.3)
\psline[linewidth=1pt](7.5,2.8)(7.5,2.2)
\end{pspicture}
\caption{The Gaussian broadcast wiretap channel with asymmetric side information.}
\end{center}
\label{figure:GaussianBroadcastWiretapStateChannel}
\end{figure}

The Gaussian broadcast wiretap channel with asymmetric side information we consider is described by the following equations:
\begin{eqnarray}
Y_1 &=& X + S_2 + W_1\\
Y_2 &=& X + S_1 + W_2\\
Z &=& X +  S_1 + S_2 + W_3
\end{eqnarray}
The random variables $W_1$, $W_2$, $W_3$, $S_1$, $S_2$ are Gaussian with mean 0 and variance $N_1$, $N_2$, $N_3$, $Q_1$, $Q_2$.
The channel states $S_1$ and $S_2$ are correlated following the parameter $\rho =\frac{\E[S_1S_2]}{\sqrt{Q_1Q_2}}$. The channel input $X$ must satisfy the constraint:
\begin{eqnarray}
\E [X^2] \leq P
\end{eqnarray}
Without loss of generality, we suppose that $N_1\geq N_2$. The channel of the first receiver is physically degradable version of the second one.
Let $\alpha_1\in \R,\alpha_2\in \R, \beta\in[0,1]$ and $\bar{\beta}=1-\beta$. Decompose $X = X_1 + X_2$ into two independent
Gaussian random variables $X_1$ and $X_2$ with mean 0 and variance $\beta P$ and $\bar{\beta} P$.
Define the following auxiliary random variables:
\begin{eqnarray*}
U_1 = X_1 + \alpha_1 S_2 &\sim& \mc{N}(0,\beta P + \alpha_1^2 Q_2)\\
U_2 = X_2 + \alpha_2 (S_1 + X_1) &\sim& \mc{N}(0,\bar{\beta} P + \alpha_2^2( Q_1 + \beta P))
\end{eqnarray*}

\begin{figure}[h!]
	\centering
		\includegraphics[width=0.5\textwidth]{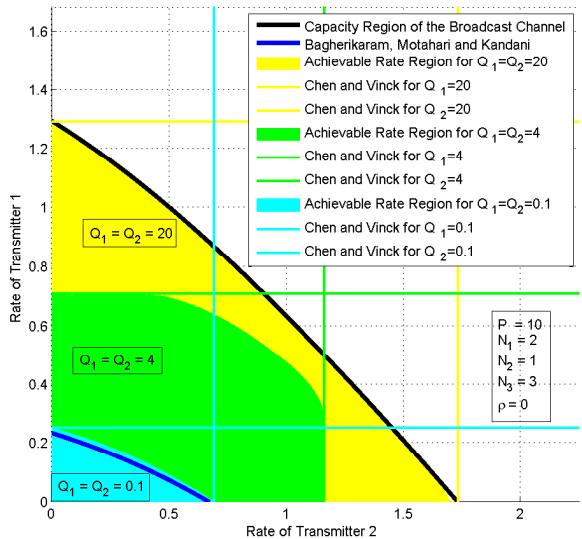}
	\caption{Rate region for the correlation parameter $\rho=0$ and different values of $Q_1$ and $Q_2$.}
\label{fig:AchievableRegion20110601}
\end{figure}

Numerical simulations (Fig. \ref{fig:AchievableRegion20110601}) illustrate the achievable rate region comparing to the previous results in \cite{Steinberg2005a},  \cite{Bagherikaram2008} and \cite{ChenVinck08}.
In Fig. \ref{fig:AchievableRegion20110601}, we compare the achievable rate region for different values $Q_1$ and $Q_2$ of the variance of the side information $S_1$, $S_2$ and for the correlation parameter $\rho=0$. When the variance of the side information is low $(Q_1=Q_2=0.1)$, the rate region (in blue) is close to the one of \cite{Bagherikaram2008}. Whereas for high variance of the side information $(Q_1=Q_2=20)$, the rate region (in yellow) is close to the capacity region for the broadcast channel of  \cite{Steinberg2005a}. High variances $Q_1$ and $Q_2$ for the side information are sufficient to compensate for the presence of an eavesdropper in the network.

\subsection{Having the side information at the decoder as well allows to enlarge the secrecy rate}\label{subsection:SideDecoder}

Often, when already available at the encoder, the knowledge of the side information at the decoder does not increase the transmission rate \cite{costa-it-1983}\cite{Steinberg2005a}.
However, this is not true when considering channels with security constraints.
We provide a special case of our channel model for which the knowledge of the side information at the decoder strictly increases the achievable rate.

\begin{figure}[!ht]
\begin{center}
\psset{xunit=0.5cm,yunit=0.5cm}
\begin{pspicture}(-1,0)(12,5)
\rput[u](1.5,0){$\mc{C}$}
\rput[u](10,2.5){$\mc{D}_1$}
\rput[u](10,0){$\mc{E}$}
\rput[u](-0.4,0.5){$m$}
\rput[u](2.5,0.4){$X$}
\rput[u](8.5,2.9){$Y_1$}
\rput[u](8.8,0.4){$Z$}
\rput[u](0.5,4.5){$S_1$}
\rput[u](11.5,3){$\hat{m}$}
\rput[d](7.7,0.7){$W_1$}
\rput[u](6.75,1.6){$W_3$}
\psframe(1,-0.5)(2,0.5)
\psframe(9.5,2)(10.5,3)
\psframe(9.5,-0.5)(10.5,0.5)
\pscircle(7.5,2.5){0.15}
\pscircle(6.75,0){0.15}
\pscircle(5,2.5){0.15}
\pscircle(5,0){0.15}
\psline[linewidth=1pt]{->}(10.5,2.5)(13,2.5)
\psline[linewidth=1pt]{->}(-1,0)(1,0)
\psline[linewidth=1pt](2,0)(3,0)
\psline[linewidth=1pt](2.75,0)(4,2.5)
\psline[linewidth=1pt]{->}(4,2.5)(4.7,2.5)
\psline[linewidth=1pt]{->}(4.7,2.5)(7.2,2.5)
\psline[linewidth=1pt]{->}(7.2,2.5)(9.5,2.5)
\psline[linewidth=1pt]{->}(1.5,4.5)(1.5,0.5)
\psline[linewidth=1pt](1,4.5)(10,4.5)
\psline[linewidth=1pt]{->}(5,4.5)(5,2.8)
\psline[linewidth=1pt]{->}(5,2.8)(5,0.3)
\psline[linewidth=1pt](5,0.3)(5,-0.3)
\psline[linewidth=1pt]{->}(10,4.5)(10,3)
\psline[linewidth=1pt]{->}(3,0)(4.7,0)
\psline[linewidth=1pt](4.7,0)(5.7,0)
\psline[linewidth=1pt]{->}(5.7,0)(6.45,0)
\psline[linewidth=1pt]{->}(6.45,0)(9.5,0)
\psline[linewidth=1pt]{->}(6.75,1.3)(6.75,0.3)
\psline[linewidth=1pt]{->}(7.5,1.2)(7.5,2.2)
\psline[linewidth=1pt](6.75,-0.3)(6.75,0.3)
\psline[linewidth=1pt](7.5,2.8)(7.5,2.2)
\end{pspicture}
\caption{The Gaussian wiretap channel with side information non-causally known at both the encoder and the decoder.}
\end{center}
\label{figure:GaussianWiretapStateSideDecoder}
\end{figure}
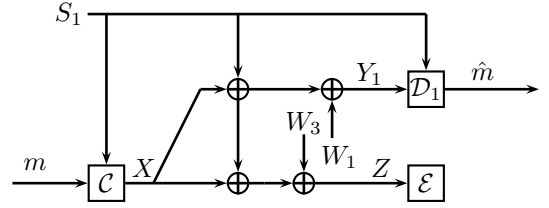

The Gaussian broadcast wiretap channel with side information at the decoder is described by the following equations:
\begin{eqnarray}
Y_1 &=& X + S_1 + W_1,\\
Z &=& X +  S_1 + W_3.
\end{eqnarray}
This channel is a special case of the model we consider here above when we remove the decoder $\mc{D}_2$ and we fix the second side information constant $\mc{S}_2=\emptyset$.
The side information $S_1$ is non-causally known at the decoder. The random variables $W_1$, $W_3$, $S_1$, are gaussian with mean 0 and variance $N_1$, $N_3$, $Q_1$.
The channel input $X$ must satisfies the constraint:
\begin{eqnarray}
\E [X^2] \leq P
\end{eqnarray}

\begin{theorem}
The capacity of the channel with state is achievable.
\begin{eqnarray*}\label{inequalities:Capacity}
\C &=& I(U_1;Y_1|S_1).
\end{eqnarray*}
\end{theorem}

The proof consists in replacing the random variable $U_1$ with a parameter $\alpha_1\gg 1$ in the first equation of (\ref{inequalities:maintheorem}).

\section{Min-max level for a long-run game with signals}\label{section:MinmaxLevels}

The above-referenced channel is now used to model the transmission of strategic information in a long-run
 game with signals that is, a game where a given player has a certain observation of the actions played by
 the others \cite{AumannMashler(BookRGIncInfo)95}. Therefore, in dynamic games with imperfect monitoring/observation, players observe the actions
taken by other players through channels also called ``signalling structure''. An important challenge is to
characterize the set of equilibrium utilities for a long-run game with imperfect monitoring; even in the
  case of repeated games, the problem of finding this set is still open \cite{RenaultTomala(GeneralProp)11}.
  This problem is closely related
  to the characterization of achievable rate regions for a class of channel models containing the one we
  investigate in this paper. Coding/decoding schemes designed for channels with security constraints can allow
   a group of players to correlate their sequence of plays keeping it secret from another group of players.
   Our main contribution is to point out a general methodology which can be used in many other scenarios and
   provide, for a specific example an upper bound on min-max levels. The example chosen is a four-player
   repeated game with signals, directly establishing a link with the multiuser channel studied in Sec. \ref{section:ChannelModel}.

\subsection{A repeated game with signals}\label{subsection:RGwithSignals}

A stage game is defined by a set of players $\mc{K}$, each of them having a set of actions $\mc{A}_k$ and a stage-utility function $u_k$.
In a long run game, a strategy $\tau_k=(\tau_k^t)_{1\leq t}$ of player $k\in \mc{K}$ is a sequence of functions from the sequences of signals $S_k^{\times(t-1)}$ into the mixed actions $\Delta(\mc{A}_k)$:
\begin{eqnarray}
\tau_k^t : \mc{S}_k^{\times(t-1)} &\longrightarrow& \Delta(\mc{A}_k)
\end{eqnarray}
A profile of strategies $\tau =(\tau_k)_{i\in \mc{K}}$ induces a probability distribution $\PP_{\tau}\in\Delta(\mc{A}^{\infty})$ over the sequences of actions $(a^t)_{t\geq1}$. The utility of the $n$-stage game is related to the above probability  $\PP_{\tau}$.
\begin{eqnarray}
\gamma_k^{n}(\tau) &=& \E_{\tau} \frac{1}{n}\sum_{t= 1}^n  u_k(a_1^t,\ldots,a_K^t)
\end{eqnarray}
The reader is referred to the paper of Renault and Tomala \cite{RenaultTomala(GeneralProp)11} for more details about the model of repeated games with signals.

\subsection{The min-max levels as ``punishment levels''}\label{subsection:PunishmentLevels}

The min-max level, also called ``the punishment level'', of a player measures the worst utility
level this player can be forced by the others in a long-run game. The formal problem of the
min-max levels is in the articles of Gossner and Tomala \cite{GossnerTomala06}, \cite{GossnerTomala07}.
They provide a characterization of the min-max using entropy methods. Denote $\tau_{-k}$ the
vector of strategy of all the players $\ell\neq k\in \mc{K}$ except $k\in \mc{K}$.

\textit{\begin{definition}
The uniform min-max $v_k^{\infty}$ for player $k\in \mc{K}$ is defined as follows:
\begin{itemize}
\item[$\bullet$] The players $\ell\neq k \in \mc{K}$ guarantee $v_k^{\infty}\in R$ if:
\begin{eqnarray}
&&\forall \varepsilon>0,\exists \tau_{-k},\exists N\in \N, \forall \tau_k,\forall n\geq N\\
&&\quad \gamma_k^n (\tau_k,\tau_{-k})  \leq v_k^{\infty} + \varepsilon
\end{eqnarray}
\item[$\bullet$] The player $k \in \mc{K}$ defends $v_k^{\infty}\in R$ if:
\begin{eqnarray}
&&\forall \varepsilon>0, \forall \tau_{-k},\exists \tau_k, \exists N\in \N, \forall n\geq N\\
&&\quad \gamma_k^n (\tau_k,\tau_{-k})  \geq v_k^{\infty} - \varepsilon
\end{eqnarray}
\item[$\bullet$] The uniform min-max of player $k\in \mc{K}$, if it exists, is $v_k^{\infty}\in \R$ such that players $\ell\neq k \in \mc{K}$ guarantee $v_k^{\infty}\in R$ and player $k\in \mc{K}$ defends $v_k^{\infty}\in R$.
\end{itemize}
\end{definition}}

\subsection{Upper bound on min-max levels}\label{subsection:MainResult}

We denote $\mc{A}_{123}=\mc{A}_1\times \mc{A}_2\times \mc{A}_3$ the product of actions set and $X_{123}=\prod_{k=1,2,3}\Delta(\mc{A}_k)$ the product of independent probabilities over the player's actions.

\begin{definition}
Define $\Q_1\subset \Delta(\mc{A}_1\times \mc{A}_2\times \mc{A}_3)$ the set of achievable empirical distributions,  where player $P_1$ is the encoder, such that for all $\QQ_1\in \Q_1$ there exists  a distribution,
$$\widetilde{\QQ_1}\in \Delta(\mc{U}_2\times \mc{U}_3\times \mc{A}_1\times\ldots \mc{A}_3\times \mc{S}_1\times\ldots \mc{S}_4)$$
 satisfying the two following conditions:
 \begin{itemize}
 \item[$\bullet$]the conditions on the marginals:
\begin{eqnarray*}
\sum_{u,s}\widetilde{\QQ_1}(u,a,s)&=&\QQ_1(a)\\
\widetilde{\QQ_1}(s|u,a)&=&T(s_2,s_3,s_4|a_1,a_2,a_3)
\end{eqnarray*}
 \item[$\bullet$]the information theoretical conditions:
 \begin{eqnarray*}\label{Cond:EntropyConstraints}
H_{}(A_2)&\leq & I_{ }(U_2;S_2,A_2) \\
&-& \max(I_{ }(U_2;S_4), I_{ }(U_2;A_2,A_3))\\
H_{ }(A_3)&\leq & I_{ }(U_3;S_3,A_3) \\
&-& \max(I_{ }(U_3;S_4), I_{ }(U_3;A_2,A_3))\\
H_{ }(A_2) &+&H_{ }(A_3) \leq I_{ }(U_2;S_2,A_2)\\
 &+&I_{ }(U_3;S_3,A_3)  - I_{ }(U_2;U_3)\\
  &-& \max(I_{ }(U_2,U_3;S_4), I_{ }(U_2,U_3;A_2,A_3) )
\end{eqnarray*}
\end{itemize}
Define in a similar way $\Q_2$ (resp. $\Q_3$), when player $P_2$ (resp. player $P_3$) is an encoder in the above channel model. Let $\Q_{123}$ denote the convex hull of the union of achievable distributions when one of the players is an encoder:
\begin{eqnarray*}
\Q_{123} = \text{co } [\Q_1\cup \Q_2\cup\Q_3\cup X_{123}] \subset \Delta(\mc{A}_{123})
\end{eqnarray*}
\end{definition}

\begin{theorem}\label{theo:AchievMin-max4}
Suppose that the channel transition $T$ does not depend on the actions of the fourth player:
\begin{eqnarray*}
T(s_1,s_2,s_3,s_4|a_1,a_2,a_3,a_4)&=&T(s_1,s_2,s_3,s_4|a_1,a_2,a_3),\\
&&\; \forall a_k,s_k,\; k\in \mc{K}
\end{eqnarray*}
The uniform min-max level $v_4^{\infty}$ of player $P_4$ for the repeated game with signals is upper bounded by the following quantity:
\begin{eqnarray*}
v_4^{\infty}\leq \min_{\QQ \in \Q_{123}}\max_{a_4\in \mc{A}_4} \E_{Q}u_4(a_1,a_2,a_3,a_4)=\nu
\end{eqnarray*}
\end{theorem}

\subsection{Sketch of the proof of Theorem \ref{theo:AchievMin-max4}}\label{subsection:SketchProof}
\begin{figure}[!ht]
\begin{center}
\psset{xunit=0.5cm,yunit=0.5cm}
\begin{pspicture}(-0.5,-0.5)(14,5)
\psline[linewidth=2pt](0,0)(14,0)
\psline[linewidth=2pt](0,2)(14,2)
\psline[linewidth=2pt](0,4)(14,4)
\psline[linewidth=1pt](2,-1)(2,5)
\psline[linewidth=0.5pt,linestyle=dashed ](4,-1)(4,5)
\psline[linewidth=0.5pt,linestyle=dashed ](6,-1)(6,5)
\psline[linewidth=0.5pt,linestyle=dashed ](8,-1)(8,5)
\psline[linewidth=0.5pt,linestyle=dashed ](10,-1)(10,5)
\psline[linewidth=1pt](13,-1)(13,5)
\rput[u](-0.3,0){$a_3^n$}
\rput[u](-0.3,2){$a_2^n$}
\rput[u](-0.3,4){$a_1^n$}
\rput[u](2,5.4){$N_j$}
\rput[u](13,5.4){$N_{j+1}$}
\rput[u](3,4.8){$N_j^0$}
\rput[u](5,4.8){$N_j^1$}
\rput[u](7,4.8){$N_j^2$}
\pspolygon[fillstyle=vlines*](2,3.8)(2,4.2)(4,4.2)(4,3.8)
\pspolygon[fillstyle=vlines*](4,1.8)(4,2.2)(6,2.2)(6,1.8)
\pspolygon[fillstyle=vlines*](4,-0.2)(4,0.2)(6,0.2)(6,-0.2)
\psline[linewidth=1.5pt,linecolor=red  ]{->}(3,4.1)(5,1.85)
\psline[linewidth=1.5pt,linecolor=red ]{->}(3,4.1)(5,-0.15)
\end{pspicture}
\label{figure:GameCourse}
\caption{During the sub-block $N_j^0$ player $P_1$ wants players $P_2$ and $P_3$ to play certain actions during the sub-block of stages $N_j^{1}$. It can be noticed that the knowledge of the sequence of future realizations of the channel state (non-causal side information) at the encoder is therefore fully justified from a game theoretical point of view.
}
\end{center}
\end{figure}
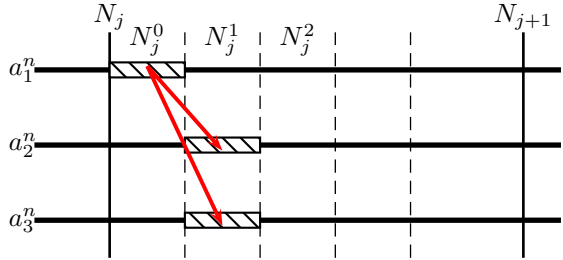

We have proven that the coding scheme described in the previous section is optimal for the players in order to guarantee the value $\nu \in \R$.
Face to the above strategy for players $P_1$, $P_2$ and $P_3$, every strategy $\tau_4$ for player $P_4$, leads to a long-run expected utility below $\nu \in \R$.
Suppose that the optimal distribution  $Q^*\in \Delta(\mc{A}_{123})$ is a convex combination:
\begin{eqnarray}
Q^* = \sum_{j=1}^J \alpha_j Q_j^* \in \Q_{123}
\end{eqnarray}
The play of players $P_1$, $P_2$ and $P_3$ is divided into $J$ blocks of stages of length $N_j$ where the players implement $Q_j^*$. Each block $\mc{N}_j$ of stages is divided into $I+1$ sub-block $\mc{N}_j^i$ where the encoding player communicate to the others, the sequence of actions they will play in the next sub-block. The recursive coding process is described in Fig. \ref{figure:GameCourse}.

\begin{figure}[!ht]
\begin{center}
\begin{tiny}
\psset{xunit=0.4cm,yunit=0.4cm}
\begin{pspicture}(-4.5,-3.7)(14,3.7)
\rput[u](1.5,0){${P}_1$}
\rput[u](10,2.5){${P}_2$}
\rput[u](10,-2.5){${P}_3$}
\rput[u](10,0){${P}_4$}
\rput[u](-0.45,0.4){$(m_1,m_2)$}
\rput[u](-4.8,0.5){$(A^{i+1}_2,A^{i+1}_3)$}
\rput[u](3.3,0.5){$\mc{A}_1^i$}
 \rput[u](5.5,0){$T$}
\rput[u](8.5,3){$S_2^i$}
\rput[u](8.5,-2){$S_3^i$}
\rput[u](8.5,0.5){$S_4^i$}
\rput[u](0.5,3.5){$\mc{A}_2^i$}
 \rput[u](0.5,-3.6){$\mc{A}_3^i$}
\rput[u](11.5,2.9){$\hat{m}_1$}
 \rput[u](11.5,-2.1){$\hat{m}_2$}
 \rput[u](14.5,3){$\hat{A}^{i+1}_2$}
 \rput[u](14.5,-2){$\hat{A}^{i+1}_3$}
 \rput[u](-2.5,0){${P}_1$}
\rput[u](13,2.5){${P}_2$}
\rput[u](13,-2.5){${P}_3$}
\psframe(-3,-0.5)(-2,0.5)
\psframe(1,-0.5)(2,0.5)
\psframe(9.5,2)(10.5,3)
\psframe(9.5,-3)(10.5,-2)
\psframe(9.5,-0.5)(10.5,0.5)
\psframe(12.5,2)(13.5,3)
\psframe(12.5,-3)(13.5,-2)
\pscircle(5.5,0){0.28}
\psline[linewidth=1pt]{->}(10.5,2.5)(12.5,2.5)
\psline[linewidth=1pt]{->}(10.5,-2.5)(12.5,-2.5)
\psline[linewidth=1pt]{->}(13.5,2.5)(15,2.5)
\psline[linewidth=1pt]{->}(13.5,-2.5)(15,-2.5)
\psline[linewidth=1pt]{->}(-2,0)(1,0)
\psline[linewidth=1pt]{->}(-5.5,0)(-3,0)
\psline[linewidth=1pt]{->}(2,0)(4.8,0)
\psline[linewidth=1pt](6,0.44)(8,2.5)
\psline[linewidth=1pt](6,-0.44)(8,-2.5)
\psline[linewidth=1pt]{->}(8,2.5)(9.5,2.5)
\psline[linewidth=1pt]{->}(8,-2.5)(9.5,-2.5)
\psline[linewidth=1pt]{->}(1.5,-3.5)(1.5,-0.5)
\psline[linewidth=1pt]{->}(1.5,3.5)(1.5,0.5)
\psline[linewidth=1pt](1,-3.5)(10,-3.5)
\psline[linewidth=1pt](1,3.5)(10,3.5)
\psline[linewidth=1pt]{->}(5.5,-3.5)(5.5,-0.7)
\psline[linewidth=1pt]{->}(5.5,3.5)(5.5,0.7)
\psline[linewidth=1pt]{->}(10,3.5)(10,3)
\psline[linewidth=1pt]{->}(10,-3.5)(10,-3)
\psline[linewidth=1pt]{->}(6.2,0)(9.5,0)
\end{pspicture}
\label{figure:JointSourceChannel}
\end{tiny}
\caption{The joint source channel coding scheme for transmitting during the sub-block of stages $\mc{N}_j^{i}$ the actions of the sub-block of stages $\mc{N}_j^{i+1}$.}
\end{center}
\end{figure}
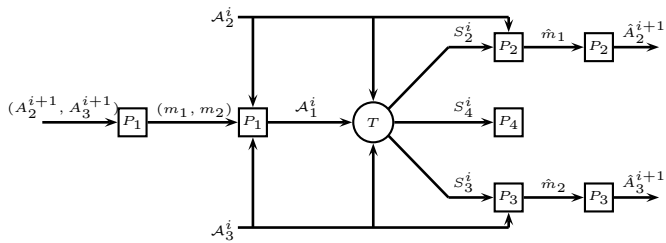

For each sub-block $i\in I$, the coding scheme consists of a concatenation of the Shannon's source coding scheme \cite{Cover-Book-91} and the channel coding scheme investigated here above. The joint source coding scheme is described in Fig. \ref{figure:JointSourceChannel} where $\mc{A}_k^i$ denotes the sequence of actions of player $P_k$ during the sub-block of stages $\mc{N}_j^i$. The entropy constraints (\ref{Cond:EntropyConstraints}) in the definition of $\Q_{123}$ insure that the sequence of actions of players can be sent over the channel and recovered with an arbitrary small error probability.

Our coding scheme guarantees that the expectation of the empirical distribution of plays $\E[\bar{Q}]$ converges to the optimal distribution $Q^*$. Second, the coding scheme guarantees  that the distribution over the signals $s_4^n$ of player $P_4$ prevents her to guess the future sequence of correlated actions of the players $P_1$, $P_2$ and $P_3$.

\section{Conclusion}\label{conclusion}

This paper investigates a generalization of the wiretap channel with two receivers and one eavesdropper
where the channel transition depends on states known non-causally and perfectly at the encoder and partially
known at both receivers. The main theorem of the paper provides an achievable rate region. Applying the
theorem to the Gaussian case allows one to make several interesting observations. In particular, two scenarios have been studied. In the first scenario, we have shown that, contrarily to \cite{costa-it-1983} and related works, having side information at the decoder in addition
to having it at the encoder is useful when security constraints come into play. Whereas this result
has been proved for the Gaussian case, further works should be necessary to study the discrete case
(e.g., by introducing more auxiliary variables to fully exploit the knowledge of the side information
at the encoder). In the second scenario, it is shown that the presence of known perturbations (namely $S_1$ and $S_2$) can enhance the secrecy rates. In fact, if those perturbations are sufficiently strong, it is even possible to obtain the same rate region as if the eavesdropper were not present. Another type of interesting result is that we show how multiuser Shannon theory can
be exploited for general games, opening a general methodology to derive communication-compatible game-theoretic such as min-max levels, feasible joint distributions or correlated strategies, etc. One the key observations
made in this paper is that source-channel theorems might play an increasing role in games
where inter-player communications is allowed.


\bibliographystyle{plain}
\bibliography{BiblioMael}

\begin{thebibliography}{10}

\bibitem{AumannMashler(BookRGIncInfo)95}
R.~J. Aumann, M.~Maschler, and R.~E. Stearns.
\newblock {\em Repeated Games with Incomplete Information}.
\newblock The MIT Press, 1995.

\bibitem{Bagherikaram2008}
G.~Bagherikaram, A.S. Motahari, and A.~K. Khandani.
\newblock Secure broadcasting : The secrecy rate region.
\newblock In {\em Proc. 46th Annual Allerton Conference on Communication,
  Control, and Computing}, pages 834--841, Sept. 2008.

\bibitem{ChenVinck08}
Y.~Chen and H.~Vinck.
\newblock Wiretap channel with side information.
\newblock {\em IEEE Transactions on Information Theory}, 54(1):395--402, 2008.

\bibitem{costa-it-1983}
M.~H.~M. Costa.
\newblock Writing on dirty paper.
\newblock {\em IEEE Transactions on Information Theory}, 29:439--441, 1983.

\bibitem{Cover-Book-91}
T.M. Cover and J.A. Thomas.
\newblock Elements of information theory.
\newblock {\em Wiley-Interscience}, 1991.

\bibitem{CsiszarKorner(BroadcastConf)78}
I.~Csisz\'{a}r and J.~K\"{o}rner.
\newblock Broadcast channels with confidential messages.
\newblock {\em IEEE Transactions on Information Theory}, 24(3):339--348, 1978.

\bibitem{CsiszarKorner(Book)81}
I.~Csisz\'{a}r and J.~K\"{o}rner.
\newblock {\em Information Theory: Coding Theorems for Discrete Memoryless
  Systems}.
\newblock 1981.

\bibitem{elgamal-it-1981}
A.~A. {El~Gamal} and E.~van~der Meulen.
\newblock A proof of {M}arton's coding theorem for the discrete memoryless
  broadcast channel.
\newblock {\em IEEE Transactions on Information Theory}, 27(1):120--122, Jan.
  1981.

\bibitem{ScutariPalomarBarbarossa09}
D.~P.~Palomar G.~Scutari and S.~Barbarossa.
\newblock The mimo iterative waterfilling algorithm.
\newblock {\em IEEE Trans. Signal Process.}, 57(5):1917–1935, May. 2009.

\bibitem{gelfand-it-1980}
S.~I. Gel'fand and M.~S. Pinsker.
\newblock Coding for channel with random parameters.
\newblock {\em Problems of Control and Inform. Theory}, 9(1):19--31, 1980.

\bibitem{GossnerTomala06}
O.~Gossner and T.~Tomala.
\newblock Empirical distributions of beliefs under imperfect observation.
\newblock {\em Mathematics of Operation Research}, 31(1):13--30, 2006.

\bibitem{GossnerTomala07}
O.~Gossner and T.~Tomala.
\newblock Secret correlation in repeated games with imperfect monitoring.
\newblock {\em Mathematics of Operation Research}, 32(2):413--424, 2007.

\bibitem{KhistiTchamkertenWornell08}
A.~Khisti, A.~Tchamkerten, and G.W. Wornell.
\newblock Secure broadcasting over fading channels.
\newblock {\em IEEE Transactions on Information Theory}, 54:2453--2469, 2008.

\bibitem{Lasaulce-Tutorial-09}
S.~Lasaulce, M.~Debbah, and E.~Altman.
\newblock Methodologies for analyzing equilibria in wireless games.
\newblock {\em IEEE Signal Processing Magazine, Special issue on Game Theory
  for Signal Processing}, Sep. 2009.

\bibitem{marton-it-1979}
K.~Marton.
\newblock A coding theorem for the discrete memoryless broadcast channel.
\newblock {\em IEEE Transactions on Information Theory}, 25:306--311, Mar.
  1979.

\bibitem{RenaultTomala(GeneralProp)11}
J.~Renault and T.~Tomala.
\newblock General properties of long-run supergames.
\newblock {\em Dynamic Games and Applications}, 1(2):319--350, 2011.

\bibitem{shannon-bell-1948}
C.~E. Shannon.
\newblock A mathematical theory of communication.
\newblock {\em Bell System Technical Journal}, 27:379--423, 1948.

\bibitem{Shannon(secrecy)1949}
C.~E. Shannon.
\newblock Communication theory of secrecy systems.
\newblock {\em Bell System Technical Journal}, 28:656--715, 1949.

\bibitem{Steinberg2005a}
Y.~Steinberg and S.~Shamai.
\newblock Achievable rates for the broadcast channel with states known at the
  transmitter.
\newblock In {\em Proc. International Symposium on Information Theory ISIT
  2005}, pages 2184--2188, 4--9 Sept. 2005.

\bibitem{YuGinisCioffi2002}
G.~Ginis W.~Yu and J.~M. Cioffi.
\newblock Distributed multiuser power control for digital subscriber lines.
\newblock {\em IEEE J. Sel. Areas Commun.}, 20(5):1105–1115, May. 2002.

\bibitem{Wyner(Wiretap)1975}
A.~D. Wyner.
\newblock The wire-tap channel.
\newblock {\em The Bell System Technical Journal}, 54(8):1355--1387, 1975.

\end{thebibliography}

\end{document}